\documentclass[twocolumn,trackchanges]{aastex62} 

\submitjournal{ApJ}

\shorttitle{Probing $G_{\rm eff}(z)$ with optimal redshift weights}
\shortauthors{Li \& Zhao}

\def\be{\begin{equation}}
\def\ee{\end{equation}}
\def\ba{\begin{eqnarray}}
\def\ea{\end{eqnarray}}

\def\ie{{\frenchspacing\it i.e.}}
\def\eg{{\frenchspacing\it e.g.}}

\def\nn{\nonumber}

\def\d{{\rm d}}

\def\nunit{{h^3 \ {\rm Mpc}^{-3}}}
\def\f{\frac}
\def\l{\left}
\def\r{\right}

\def\lnf{{\rm ln} f}

\def\lnP{{\rm ln} P}
\def\lnk{{\rm ln} k}
\def\lnPm{{\rm ln} P_0}
\def\lnPq{{\rm ln} P_2}
\def\lnPh{{\rm ln} P_4}

\def\b{\beta}
\def\bT{b_{\rm T}}
\def\hoMpc{h^{-1} {\rm Mpc}}

\newcommand{\OM}{\Omega_{\rm M}}
\def\G{G_{\rm eff}}
\def\W{\bf W}

\def\p{\partial}
\def\ab{\alpha_{\bot}}
\def\ap{\alpha_{\parallel}}
\def\cf{\chi_{\rm f}}
\def\zp{z_{\rm p}}
\def\fzp{f_{\zp}}
\def\Hf{H_{\rm f}}
\def\pppk{\f{\p\lnP_{\rm m}}{\p\lnk}}

\begin{document}

\title{Probing the time variation of the effective Newton's constant with optimal redshift weights}

\author{Jian Li}\thanks{E-mail: \url{jli@nao.cas.cn}}
\affiliation{National Astronomical Observatories, Chinese Academy of Science, Beijing, 100101, P.R.China}

\affiliation{University of Chinese Academy of Sciences, Beijing, 100049, P.R.China}

\author[0000-0003-4726-6714]{Gong-Bo Zhao}\thanks{E-mail: \url{gbzhao@nao.cas.cn}}\affiliation{National Astronomical Observatories, Chinese Academy of Science, Beijing, 100101, P.R.China}
\affiliation{School of Astronomy and Space Science, University of Chinese Academy of Sciences, Beijing, 100049, P.R.China}
\affiliation{Institute of Cosmology and Gravitation, University of Portsmouth, Portsmouth, PO1 3FX, UK}

\begin{abstract}
We propose a new method for probing the time variation of the effective Newton's constant $G_{\rm eff}$, based on the optimal redshift weighting scheme, and demonstrate the efficacy using the DESI galaxy spectroscopic survey. We find that with the optimal redshift weights, the evolution of $\G(z)$ can be significantly better measured: the uncertainty of $\G(z)$ can be reduced by a factor of $2.2\sim12.8$ using the DESI BGS sample at $z \lesssim0.45$, and by a factor of $1.3\sim4.4$ using the DESI ELG sample covering $0.65\lesssim z\lesssim1.65$.
\end{abstract}


\section{Introduction} \label{sec:intro}

The cosmic acceleration discovered in 1998 \citep{Riess,Perlmutter} may imply that an unknown energy component with a negative pressure, dubbed dark energy (DE), could contribute significantly to the Universe at recent epoch, or that Einstein's general relatively (GR) needs to be extended or altered on cosmological scales. 

As dark energy does not cluster below the horizon scale in most dark energy models, the nature of DE is probed by the background expansion history of the Universe (see \citealt{DEreview1,DEreview2} for reviews on DE), using the supernovae type Ia (SN Ia) as ``standard candles", and/or the baryonic acoustic oscillations (BAO) scale as a ``standard ruler" of the Universe \citep{BAO1,BAO2}. On the other hand, the scenario of modified gravity (MG), which is an alternative to DE as a possible solution to the cosmic acceleration problem, has been investigating extensively from sub-galactic to cosmological scales both theoretically and observationally (see \citealt{MGreview1,MGreview2,MGreview3} for recent reviews).

In GR, the Newton's constant, $G_{\rm eff}$, and the gravitational slip, $\eta$, which is the ratio between two gravitational potentials, are both unity, but they can be functions of cosmic time and scales in general modified gravity scenarios \footnote{Note that $G_{\rm eff}$ and $\eta$ defined here are the same quantities as $\mu$ and $\gamma$, respectively, as defined in the {\tt MGCAMB} paper and code \citep{MGCAMB2}.}. Hence a deviation of either $G_{\rm eff}$ or $\eta$ from unity evidenced by observations can be a smoking gun of MG.

While $\eta$ is most efficiently probed by the weak gravitational lensing (WL) with galaxy imaging surveys, or by the integrated Sachs-Wolfe (ISW) \citep{ISW} effect probed by the cosmic microwave background (CMB) experiments, $G_{\rm eff}$ is best measured by galaxy spectroscopic surveys through the redshift-space distortions (RSD) \citep{RSD1,RSD2}. As $G_{\rm eff}$ determines the growth of cosmic structures on sub-horizon scales, it is usually better measured than $\eta$, according to a principal component analysis (PCA) on the general $G_{\rm eff}$ and $\eta$ functions \citep{MGPCA1,MGPCA2, MGPCA3, MGPCA4, MGPCA5,LZ18}. In this work, we focus on probing the time evolution of $G_{\rm eff}$ using redshift surveys.

To probe the temporal evolution of $G_{\rm eff}$, we need tomographic information of the clustering of galaxies on the past lightcone, which can be extracted from galaxy surveys using overlapping redshift slices \citep{Zhaotomo16,Wangtomo16,Wang17RSD,Zheng18}. A more computationally efficient method, which is based on the optimal redshift weighting scheme, has been recently developed and implemented for the measurement of BAO and RSD \citep{ZPW,Zhu18,RR18,DD18,Zhao18}. {The basic idea of this method is the following. The key information to constrain the concerning cosmological parameters, say the BAO and RSD parameters, is actually combinations of the galaxy power spectra at various redshifts. These combinations can be obtained by assembling the power spectra measured at a large number of redshifts. However, this is inefficient and subject to systematics. Alternatively, one can assign a weight to each individual galaxy in the catalog according to its redshift $z$, and then measure the power spectrum of the $z$-weighted galaxy sample. Theoretically, the $z$-weights can be optimized so that the information content, which is relevant for the parameters concerned, of the power spectra measured from the $z$-weighted sample is the same as that measured from a large number of redshift slices. This is essentially an optimal data compression procedure proposed in \cite{Fisher,Heavens2000} (see Sec. \ref{sec:method} for more details.). As the optimal $z$-weights only need to be computed once using a fiducial cosmology, this method is much more efficient. Moreover, the systematics can be better controlled, as the analysis is performed on the entire galaxy catalog, instead of on a large number of redshift slices, which contain far less number of galaxies in each slice.} 

In this work, we propose to measure the evolution of $G_{\rm eff}$ using the optimal redshift weighting method, and demonstrate the efficacy using a worked example of the Dark Energy Spectroscopic Instrument (DESI) survey \citep{DESI} \footnote{\url{https://www.desi.lbl.gov/}}.

We present the methodology in the next section, followed by a demonstration using the DESI galaxy survey in Section \ref{sec:result}, before conclusion and discussions in Section \ref{sec:conclusion}.

\begin{figure*}
\centering
{\includegraphics[scale=0.45]{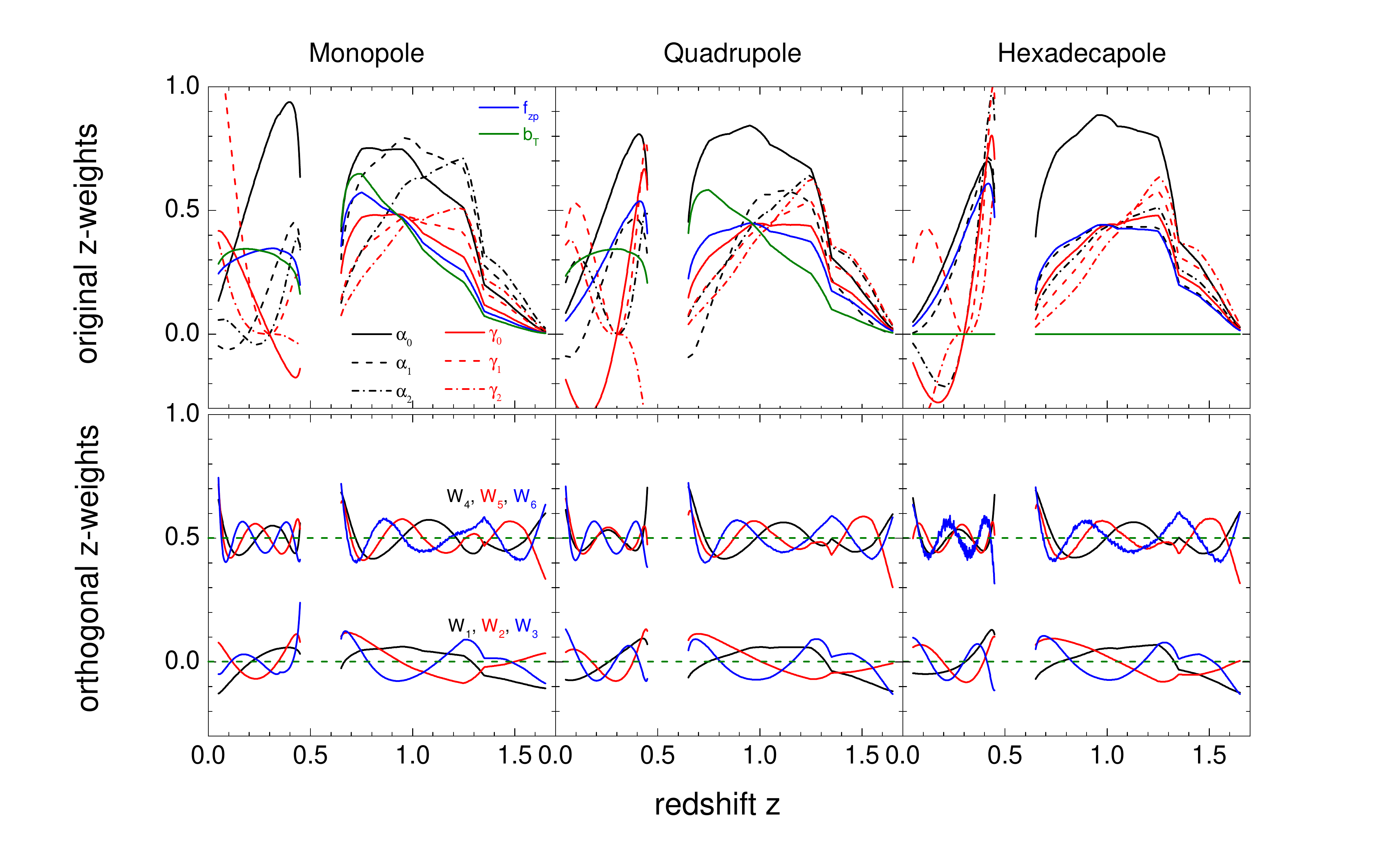}}
\caption{Upper panels: the original optimal redshift weights for the eight parameters shown in Eq (\ref{eq:param}) for the power spectrum monopole (left), quadrupole (middle) and hexadecapole (right); Lower panels: the orthogonal redshift weights derived from a SVD analysis on the original weights. The fourth to sixth redshift weights are offset by $+0.5$ for visualization.}   
\label{fig:weight}
\end{figure*}

\begin{figure*}
\centering
{\includegraphics[scale=0.4]{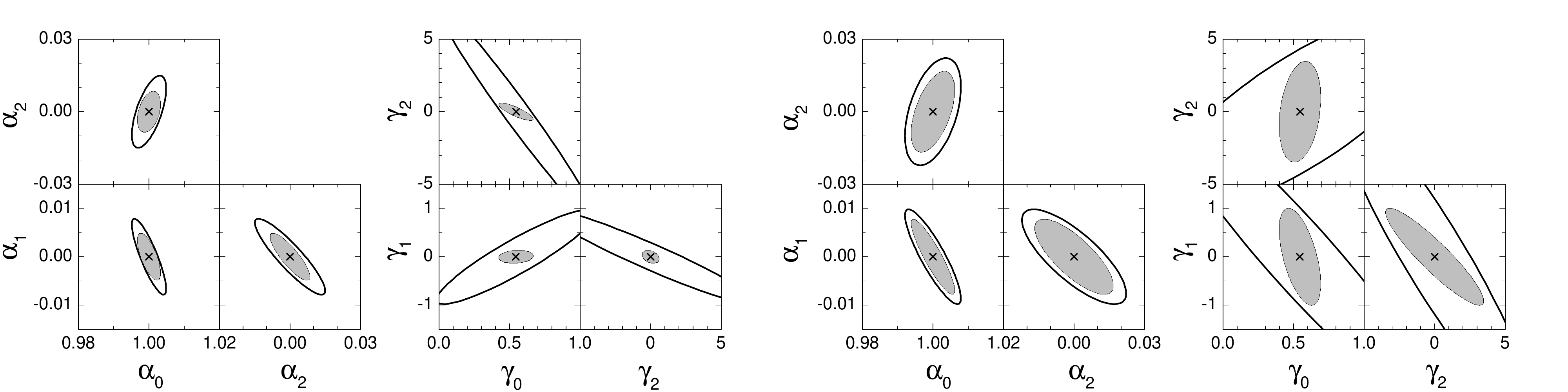}}
\caption{The 68\% CL contour plots for the $\alpha$ and $\gamma$ parameters. The inner (filled) and outer (unfilled) contours are derived with and without the optimal redshift weights applied. The left two and right two panels are derived from the DESI BGS and ELG samples respectively. The black crosses in the centre denote the fiducial values of the parameters.}
\label{fig:contour}
\end{figure*}

\begin{figure}
\centering
{\includegraphics[scale=0.3]{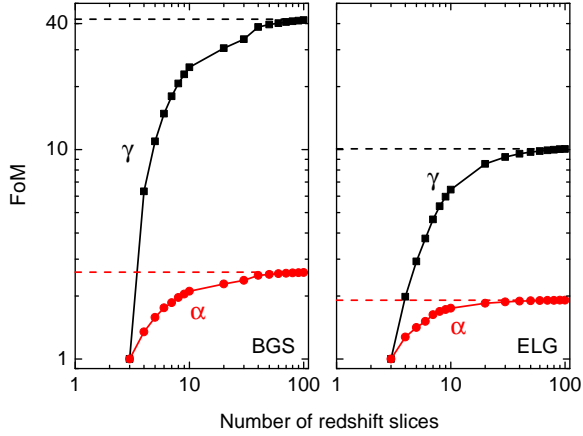}}
\caption{{Symbols connected by solid lines:} the FoM of the $\gamma$ parameters (black squares) and $\alpha$ parameters (red circles) {derived using the conventional method,} as a function of redshift slices for the BGS (left panel) and ELG samples (right). The horizontal dashed lines show the FoM with the optimal redshift weights applied. In all cases, the FoM is normalized using that with three redshift slices.}   
\label{fig:FoM}
\end{figure}

\begin{figure*}
\centering
{\includegraphics[scale=0.4]{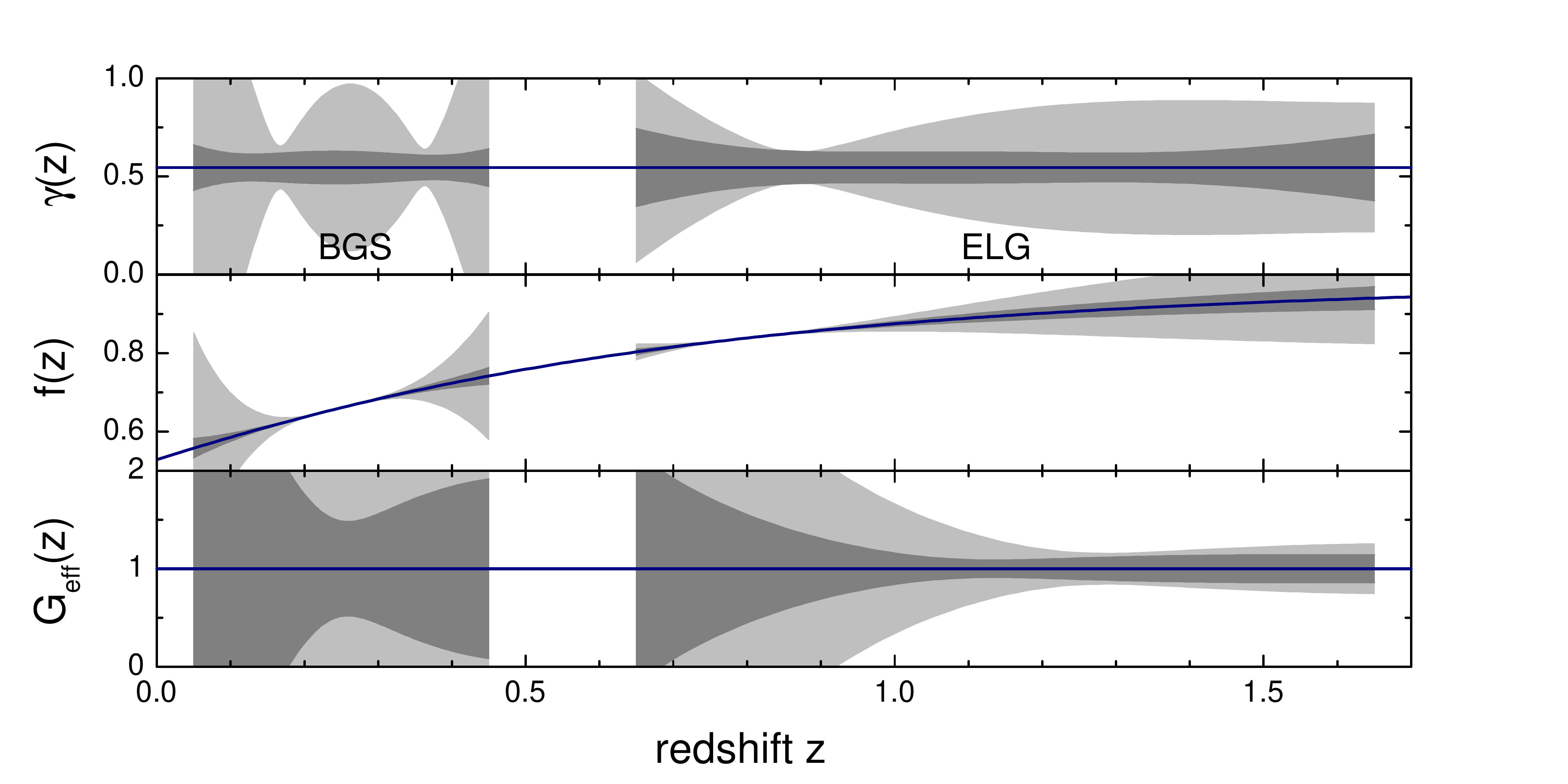}}
\caption{The forecasted 68\% CL uncertainty of $\gamma(z)$ (upper panel), $f(z)$ (middle) and $G_{\rm eff}(z)$ (lower) derived with (inner dark-shaded) and without (outer light-shaded) the optimal redshift weights. The bands below redshift $0.45$ and beyond redshift $0.65$ are derived from the DESI BGS and ELG samples respectively. The blue curves in the centre denote the fiducial model.}
\label{fig:Geff}
\end{figure*}

\section{Methodology} \label{sec:method}

In this section, we develop the methodology to measure the time variation of $\G$, after a brief review of the optimal redshift weighting method.

The optimal redshift weighting technique is essentially a data compression scheme based on the Karhunen-Lo\`eve (K-L) compression method \citep{Fisher,Heavens2000}. To illustrate the idea, let us assume that we use $N_p$ parameters to parameterize the galaxy power spectra multipoles in redshift space, which are measured at $N_z$ redshifts and at $N_k$ wavenumbers. The power spectrum {data vector} ${\bf P}$ is defined as, 
\ba     {\bf P} &\equiv& \left[{\bf P}_{z}(z_1),{\bf P}_{z}(z_2), ..., {\bf P}_{z}(z_{N_z})\right]^T, \nn\\
              {\bf P}_{z}(z_i)& \equiv&\left[{\bf P}_{0,z}(z_i),{\bf P}_{2,z}(z_i),...,{\bf P}_{2N_{\ell},z}(z_i)\right]^T, \nn\\
 \label{eq:Plz}             {\bf P}_{\ell,z}(z_i)& \equiv&\left[P_{\ell}(k_1,z_i),P_{\ell}(k_2,z_i),...,P_{\ell}(k_{N_k},z_i)\right]^T.
 \ea {where $\ell,k$ and $z$ denote the order of Legendre polynomial, the wavenumber and redshift, respectively ({so ${\bf P}$ is a column vector with $N_z\times N_{\ell}\times N_k$ rows).}}
 
The Fisher information matrix ${\bf F}$ using observables ${\bf P}$ is then, \be\label{eq:FF} {\bf F}={\bf D}^T {\bf C}^{-1}  {\bf D}, \ee where ${\bf C}$ is the data covariance matrix, and the matrix ${\bf D}$ stores the derivative of ${\bf P}$ with respect to the parameters.

It can be proved that if the data vector and covariance are compressed by the optimal weighting matrix $\W$, \ie, \ba\label{eq:W} \W={\bf C}^{-1} {\bf D}; \ \ {\bf P}_{\rm w} \equiv {\W}^T {\bf P}; \ \ {\bf C}_{\rm w}\equiv{\bf W}^T~{\bf C}~{\bf W}.\ea Then the compression is lossless, as the Fisher matrix using the compressed observables {(\ie, ${\bf P}_{\rm w}$ and ${\bf C}_{\rm w}$ are used as the data vector and data covariance matrix, respectively)}, has identical information as the uncompressed one, \ie, \ba\label{eq:FwF} {\bf F}_{\rm w}\equiv {\bf D}_{\rm w}^T {\bf C}_{\rm w}^{-1}  {\bf D}_{\rm w}={\bf F}.\ea We refer the readers to \citet{Zhao18} for a proof of Eq. (\ref{eq:FwF}), and to \citet{ZPW,Zhu18,RR18,DD18,Zhao18} for implications of this method for BAO and RSD measurements using mock or actual galaxy catalogs \footnote{Note that the weight ${\bf W}$ is optimal in $z$, although it has an index of $\ell$, and a dependence on $k$. As shown in Eq (\ref{eq:Plz}), we keep the $\ell$-dependence when evaluating the weights, but drop the $k$-dependence as it is rather weak (see discussions at the end of Sec. \ref{sec:method}).}.

In this work, we derive the optimal redshift weights for the effective Newton's constant $\G$, which is a function of redshift $z$ in general, and quantify how much improvement can be obtained with the redshift weights applied, using specifications of DESI to demonstrate \footnote{The optimal redshift weight shown in Eq (\ref{eq:W}) is estimated using a theoretical template Eq (\ref{eq:Pg}) based on a fiducial model. As the theoretically modeled power spectra may be different from the observed ones, the weights may not be exactly optimal, which may be subjective to information loss when using the weighted observables. But as long as the data vector and theoretical prediction are weighted in the same way, it is not expected to give rise to theoretical systematics, although it needs to be checked quantitatively. This is beyond the scope of this paper, and is left for a future study.}.

As we focus on galaxy redshift surveys for this study, we parametrize the time evolution of the power spectrum multipoles as follows,

\ba\label{eq:Pg} &&P_{\rm g}(k,\mu,z) = \frac{1}{\alpha_{\bot}^2\alpha_{\parallel}}\left[b+f\mu'^2\right]^2 P_{\rm m}(k',z)D_{\rm FoG}(k',\mu',z), \nonumber \\
&&D_{\rm FoG}(k',\mu',z)\equiv{\rm Exp}\left[-\frac{1}{2}\left(k_{\bot}'^{2}\Sigma_{\bot}^2 + k_{\parallel}'^2\Sigma_{\parallel}^2 \right)\right], \ea where $b$ and $f$ are the linear bias and logarithmic growth rate respectively, and $P_{\rm m}$ denotes the linear matter power spectrum {evaluated using {\tt CAMB} \citep{CAMB}}. The prime on $k$ and $\mu$ denotes the mode distorted by the Alcock-Paczynski (AP) effect \citep{AP}, and $\alpha_{\bot}$ and $\alpha_{\parallel}$ are the dilation parameters for BAO distances,\footnote{Throughout the paper, the super- or sub-script `fid' denotes the quantity for the fiducial cosmology.},
\ba\label{eq:alpha} \alpha_{\bot}=\f{D_A}{D_A^{\rm fid}}\f{r_s^{\rm fid}}{r_s}; \ \  \alpha_{\parallel}=\f{H^{\rm fid}}{H}\f{r_s^{\rm fid}}{r_s}. \ea The damping term $D_{\rm FoG}$ quantifies the Fingers of God (FoG) effect, in which quantities $k_{\bot}\equiv k\sqrt{1-\mu^2}$ and $k_{\parallel}\equiv k\mu$ represent the transverse and radial wavenumber respectively, and $\Sigma_{\bot}=\Sigma_0 G, \ \Sigma_{\parallel}=\Sigma_0 G(1+f)$, where $\Sigma_0$ is a constant calibrated using simulations, and $G$ and $f$ denote the linear growth function and the logarithmic growth rate respectively \citep{SE07}. The $\ell$th power spectrum multipole is an integral of $P_{\rm g}(k,\mu, z)$ in Eq (\ref{eq:Pg}), weighted by the Legendre polynomial $L_{\ell}(\mu)$, over $\mu$, \ie,
\ba\label{eq:Pell} P_{\ell}(k,z) = \int \d\mu P_{\rm g}(k,\mu, z) L_{\ell}(\mu).\ea

Note that in Eq (\ref{eq:Pg}), $\alpha_{\bot}, \alpha_{\parallel}, b, f$ are all functions of redshift\footnote{We drop the $z$-dependence of $\alpha_{\bot}, \alpha_{\parallel}, b, f$ in Eq (\ref{eq:Pg}) for brevity.}, and we parametrize the time evolution of these functions using the following forms, \\ \\
\noindent {\bf BAO:}
\ba 
\ab(z)& =& \alpha_0\left(1+\alpha_1 x+\frac{1}{2}\alpha_2 x^2\right), \nn \\
      \ap(z)& =& \alpha_0\left[1+\alpha_1 x+\frac{1}{2}\alpha_2 x^2+(\alpha_1+\alpha_2 x)(1+x)\right], \nn\\
 \label{eq:BAO}  x&\equiv&\f{\cf(z)}{\cf(\zp)}-1, \ea where $\cf(z)$ is the comoving distance at redshift $z$, evaluated in the fiducial cosmology, which is taken to be a $\Lambda$CDM model favored by the Planck 2018 measurement \citep{PLC18}, and $\zp$ is the pivot redshift. This is the parametrization proposed in \citet{ZPW} for the BAO, and it was shown to be an accurate approximation for general cosmologies within a wide range of redshifts. \\ \\
\noindent{\bf RSD:}    
\ba\label{eq:f} f(z)&=& f_{\zp}\l(\f{1+z}{1+\zp}\r)^{3\gamma(z)}\l[\f{\ap(z)}{\ap(\zp)}\f{\Hf(\zp)}{\Hf(z)}\r]^{2\gamma(z)},\nn\\
\label{eq:RSD}   \gamma(z) &=& \gamma_0+\gamma_1x+\frac{1}{2}\gamma_2 x^2,
\ea The parametrization of the logarithmic growth rate is essentially $f(z)=\Omega_{\rm m}^{\gamma(z)}(z)$, proposed in \cite{gammaL}, with $\Omega_{\rm m}(z)$ eliminated using the definition of $\ap$ in Eq (\ref{eq:alpha}), and the definition of $f_{\zp}\equiv f(\zp)$. We parametrize the evolution of $\gamma(z)$ using a similar expansion in $x$ as that for $\ab(z)$, which is sufficiently general to cover a wide range of $\gamma$ functions.

{The effective Newton's constant $\G(z)$ can be reconstructed from $\gamma(z)$ and $f(z)$ using \citep{MGParaLP} \footnote{Again, we drop the $z$-dependence of functions $\G,f,\OM$ and $\gamma$ here for brevity.},
\ba\label{eq:G} \G &=&\frac{2f}{3\OM}\left[f+2-3\gamma+3\left(\gamma-\frac{1}{2}\right)\OM+\gamma'{\rm ln} \ \OM\right], \nn \\ \ea where the prime denotes the derivative with respect to ${\rm ln} \ a$, and $\OM=f^{1/\gamma}$, with $f$ and $\gamma$ given by Eq (\ref{eq:f}).} \\ 

\noindent{\bf Bias:} \\ \\ We assume that the evolution of linear bias is inversely proportional to the linear growth function $D(z)$, which is normalized to unity at $z=0$ \citep{DESI}, \ie, \ba\label{eq:bias} b(z)&=&b_{\rm T}/D(z). \ea 
 
The free parameters in this study are summarized in the set ${\bf \Theta}$,  \ba\label{eq:param} {\bf \Theta} &=& \l\{\alpha_0,\alpha_1,\alpha_2,\gamma_0,\gamma_1,\gamma_2,f_{\zp},b_{\rm T}\r\}, \nn\\
&=& \l.\l\{1,0,0,6/11,0,0,\Omega_{\rm m, fid}^{6/11}(\zp),b_{\rm T}\r\}\r|_{\rm fid},
\ea where $\Omega_{\rm m, fid}(\zp)=\f{\Omega_{\rm m,0}(1+\zp)^3}{\Omega_{\rm m,0}(1+\zp)^3+(1-\Omega_{\rm m,0})}$, and $b_{\rm T}$ depends on the kind of tracer T \footnote{We assume that $\sigma_{8,0}$, the root mean square density variance within $8 \ \hoMpc$ at redshift zero, can be well determined by external observations such as the cosmic microwave background, thus we fix $\sigma_{8,0}$ to the fiducial value throughout. It is true that $\sigma_{8,0}$ is model-dependent, \eg, deriving it from the amplitude of CMB, say, $A_s$, a theoretical model is needed. But for this analysis, as we only need to evaluate the weights at a fiducial model which is taken to be the best-fit $\Lambda$CDM model constrained derived from Planck, and the uncertainty of $\sigma_{8,0}$ is much less than that of other parameters, we fix it to the value favored by Planck, which is a reasonable choice to avoid the degeneracy between $\sigma_{8,0}$ and the galaxy bias.}. 

In this work, we consider the Bright Galaxy Survey (BGS) and Emission Line Galaxies (ELG) samples of DESI observed across 14,000 square degrees of the sky, covering redshift ranges of $z\lesssim0.45$ and $0.65\lesssim z\lesssim1.65$ respectively. The BGS and ELG samples consist of $17$ million and $9.8$ million galaxies, with the maximal number density reaching $0.04$ and $0.001$ $\nunit$ respectively. We assume $b_{\rm BGS}=1.34$ and $b_{\rm ELG}=0.84$. {For the Fisher matrix analysis, we use monopole, quadrupole and hexadecapole of the galaxy power spectra, which are binned uniformly in $k$ from $0$ to $0.3 \ \hoMpc$ ($30 \ k$ bins are used for this analysis).} For more details of the target selection of these DESI tracers, we refer to \cite{DESI}.

As demonstrated in \cite{ZPW}, the pivot redshift $\zp${, which is used to define the $x$ variable in Eq (\ref{eq:BAO}),}  is a meta-parameter that chosen to be close to the centre of the redshift range of the galaxy sample concerned, to yield the best precision of the parameterizations Eqs (\ref{eq:BAO}) and (\ref{eq:f}) for a range of cosmologies. In this work, we choose $\zp=0.23$ and $\zp=0.80$ for the DESI BGS and ELG samples respectively, and we have checked that with this choice, the area of the error band of the reconstructed $\gamma(z)$ gets minimized for each of the tracers, with the redshift weights applied \footnote{Note that this is just one arbitrary choice of $\zp$. Actually as long as $\zp$ is close to the centre of the redshift range of the galaxy sample, parametrizations Eqs (\ref{eq:BAO}) and (\ref{eq:f}) are sufficiently accurate for a range of cosmologies. For example, different $\zp$ values can be used to minimize the uncertainty of $\gamma(z)$ or $\G(z)$ at a given redshift.}.

The derivatives of the power spectrum multipoles with respective to parameters are given explicitly in the Appendix (Eqs \ref{eq:A1} and \ref{eq:A2}). We model the time evolution of the data covariance matrix {\bf C} using an analytic method \citep{TNS} as follows, which has been validated using mock galaxies for the eBOSS quasar sample \citep{Zhao18}, \ba \label{eq:C} && {\rm C_{\ell\ell'}}(k,z)=\frac{4\pi^2}{k^2\Delta k\Delta V(z)}\frac{(2\ell+1)(2\ell'+1)}{2}  \nonumber\\
 &&\times \int_{-1}^{+1}\d\mu\mathcal{L}_{\ell}(\mu) \mathcal{L}_{\ell'}(\mu)\left[P_g(k,\mu,z)+\frac{1}{\bar{n}_g(z)}\right]^2\ea 
As the normalization of the redshift weights can be arbitrary, the amplitude of ${\rm C}$ is irrelevant, as long as the normalization is set the same for all redshifts. 

Given Eqs (\ref{eq:W}), (\ref{eq:A1}), (\ref{eq:A2}) and (\ref{eq:C}), the optimal redshift weights for our parameters can be evaluated. {In practice, we subdivide the redshift ranges of the BGSs and ELGs, which will be covered by DESI, into $100$ redshift slices for each of the tracers, and evaluate the redshift weights analytically at the center of each redshift bins. We have checked and found that it is unnecessary to further increase the number of redshift bins for computing the weights, as the $z$-weights saturate with $100$ $z$-bins \ie, the weights are sufficiently smooth with this number of redshift bins.} Note that the weights are generally functions of $k$ as well, but we numerically confirm that the scale-dependence is much weaker than the $z$-dependence on linear scales for the weights considered in this work. Thus we evaluate all the redshift weights at $k=0.05\hoMpc$ without loss of generality.

{To forecast how well the relevant parameters can be constrained with and without the optimal redshift weights, we adopt the Fisher matrix approach for both cases but using different observables, which is briefly summarized as follows.}

\begin{itemize}
\item {For constraints without the optimal $z$-weights: we use Eq (\ref{eq:FF}) to evaluate the Fisher matrix ${\bf F}$. As there are eight parameters to be constrained, as shown in Eq (\ref{eq:param}), we have to include some level of tomography in the observable, otherwise the parameters are perfectly degenerate, which results in a singular Fisher matrix. For this reason, we sub-divide each of the BGS and ELG samples into three sub-samples, with a $z$-binning uniform in redshift \footnote{We split the galaxies into three redshift slices in order to have sufficient data points to constrain the eight parameters, while keeping the $z$-tomography to a minimal level. This is for the purpose of a later comparison with the result with $z$-weights applied, which highlights how much the tomographic information can help improve the constraint.}.}
\item {For constraints with the optimal $z$-weights: we use Eq (\ref{eq:FwF}) instead to compute the Fisher matrix ${\bf F_W}$, using the $z$-weighted observables and data covariances given in Eq (\ref{eq:W}). Note that as we dropped the $k$-dependence of the weights since it is not significant, ${\bf F_W}$ does not exactly equal to ${\bf F}$ with infinite number of redshift bins.}
\end{itemize}

\section{Result} \label{sec:result}

This section is devoted to the main result of this work, including the optimal redshift weights derived from the simulated DESI BGS and ELG samples, and the projected constraint on the cosmological parameters with the redshift weights applied.

\begin{table}
\begin{center}
\begin{tabular}{c|cccccc}
\hline\hline
    & $\sigma(\alpha_0)$ &  $\sigma(\alpha_1)$ & $\sigma(\alpha_2)$ & $\sigma(\gamma_0)$ &  $\sigma(\gamma_1)$ & $\sigma(\gamma_2)$  \\
    \hline
 BGS&        $0.003$ & $0.005$ & $0.010$  & $0.37$ & $0.65$ & $4.96$\\
 BGS$_z$& $0.002$ & $0.003$ & $0.006$  & $0.08$ & $0.09$ & $0.40$\\
 ELG &       $0.005$ & $0.007$ & $0.015$  & $0.66$ & $2.14$ & $4.58$\\
 ELG$_z$& $0.004$ & $0.005$ & $0.011$  & $0.10$ & $0.67$ & $2.32$\\
\hline\hline
\end{tabular}
\end{center}
\caption{The uncertainty of the $\alpha$ and $\gamma$ parameters derived from DESI BGS and ELG samples respectively. The sub-script $z$ denotes the samples with the optimal redshift weights applied.}
\label{tab:param}
\end{table}%

The optimal redshift weights for parameters considered in this work are shown in the upper panels of Fig. \ref{fig:weight}. As shown, the weights for different parameters express a certain level of similarity, which can give rise to redundancy in the data space if we combine the weighted sample for the data analysis. This problem can be solved by finding the orthogonal weights using a singular-value decomposition (SVD), and only keeping the first few ``eigen-weights'' that are most informative, as proposed in \cite{Fisher,Zhao18}. In this work, we follow this approach, and derive eight orthogonal eigen-weights from the original redshift weights, ${\bf W} = {\bf U \Lambda V}^T $ where ${\bf W}$ is the data matrix of the original weights. The orthogonal redshift weights can be constructed by projecting ${\bf W}$ onto ${\bf V}$, whose variances are stored in the diagonal matrix ${\bf \Lambda}$. Keeping the first few eigen-weights largely reduce the redundancy with negligible information loss \footnote{Note that one can in principle apply the original $z$-weights to the catalogs and then remove the redundancy from the weighted samples afterwards. But this is less efficient, as it requires estimating data covariance for unnecessary data vectors.}.

For both DESI BGS and ELG samples, we find that keeping six eigen-weights for the monopole, quadrupole and hexadecapole each is sufficient to restore almost all the information in the original weights, and we show these weights in the lower panel of Fig. \ref{fig:weight}.

With these weights applied, we derive the constraint on the parameters listed in Eq (\ref{eq:param}) following the Fisher matrix approach detailed in Sec. \ref{sec:method}, and show the 68\% CL uncertainty of the $\alpha$ and $\gamma$ parameters in Table \ref{tab:param}, and the 68\% CL contour plots for the $\alpha$ and $\gamma$ parameters in shaded ellipses in Fig. \ref{fig:contour}. For the purpose of comparison, we perform an additional Fisher forecast without the redshift weights. Specifically, we assume that we will be able to split the BGS and ELG samples into $N_z$ redshift slices for each tracer, and perform BAO and RSD measurements at corresponding effective redshifts, which are assumed to distribute evenly in redshift. For this comparison, we assume that $N_z=3$ to get the unweighted constraints, as shown using unfilled contours in Fig. \ref{fig:contour}.

As illustrated, the improvement due to the redshift weights on the constraint of $\alpha$, especially on the $\gamma$ parameters, is significant. To quantify, we compute the Figure-of-Merit (FoM), which is the square root of the determinant of the $3\times3$ inverse covariance matrix for the $\alpha$ or $\gamma$ parameter blocks (with other parameters marginalized over), as follows,\ba\label{eq:FoM}
&&\f{{\rm FoM}_{\alpha} ({\rm BGS}_z)}{{\rm FoM}_{\alpha} ({\rm BGS})}= 2.6; \ \  \ \f{{\rm FoM}_{\alpha} ({\rm ELG}_z)}{{\rm FoM}_{\alpha} ({\rm ELG})}=1.9, \nn \\ 
&&\f{{\rm FoM}_{\gamma} ({\rm BGS}_z)}{{\rm FoM}_{\gamma} ({\rm BGS})}= 41.6; \ \  \f{{\rm FoM}_{\gamma} ({\rm ELG}_z)}{{\rm FoM}_{\gamma} ({\rm ELG})}=10.1,\ea where the subscript $z$ denotes the samples with the redshift weights applied. While the improvement in the FoM for the BAO parameters is around $2$ for both samples, it can be as large as a factor of ten, or forty for the ELG and BGS samples respectively, as shown in Fig. \ref{fig:FoM}. We also show the FoM for various choices of the number of redshift slices, and we see that the slope of the $\gamma$ curve is much deeper than the $\alpha$ curve at $N_z\lesssim40$ for both tracers, which means that the tomographic information is more important for the $\gamma$ parameters. This is largely due to the fact that the degeneracy among the $\gamma$ parameters can be broken by the tomographic information, which can be understood from Eq (\ref{eq:f}): the three $\gamma$ parameters can only be derived from the RSD measurement of $f$, with another four relevant parameters ($\alpha_0,\alpha_1,\alpha_2,\fzp$) marginalized over. Thus a tomographic RSD measurement at seven effective redshifts is a minimal requirement to constrain these seven parameters. On the other hand, the $\alpha$ parameters are easier to determine according to Eq (\ref{eq:BAO}): a measurement of $\ab$ and $\ap$ at two effective redshifts is sufficient to constrain the three $\alpha$ parameters, thus adding further tomographic information helps in a less significant way.

From the unfilled contours for $\gamma$ parameters in Fig. \ref{fig:contour}, we can see that with the BGS and ELG samples, which provide BAO and RSD measurements at lower and higher redshifts respectively, the direction of the degeneracy among the $\gamma$ parameters can be significantly different, even with an opposite sign. Thus one can expect that with the full tomographic information, the
degeneracy can be largely broken. This is actually the case: with the redshift weights, $\rho(\gamma_0,\gamma_1)$, the correlation coefficient between $\gamma_0$ and $\gamma_1$ is reduced from $0.91$ for the BGS sample to $0.14$, and  $\rho(\gamma_1,\gamma_2)$ drops from $0.78$ to $0.27$ for the ELG sample. 

The result is shown in Fig. \ref{fig:Geff}. As expected, the constraint on the functions of $\gamma,f$ and $\G$ gets significantly tightened by the redshift weights, \ie, the uncertainty of $\G$ is reduced by a factor of $2.2\sim12.8$ for the redshift range covered by the BGS sample, and by a factor of $1.3\sim4.4$ by the ELG sample.

\section{Conclusion and Discussions} \label{sec:conclusion}

Probing the time evolution of the Newton's constant $\G$ plays a key role in cosmological tests of gravity. In this work, we develop a novel method based on the optimal redshift weighting scheme to probe the temporal evolution of $\G$ using galaxy spectroscopic surveys.

We start by parametrizing the evolution of BAO, RSD and bias functions in Eqs (\ref{eq:BAO}), (\ref{eq:f}) and (\ref{eq:bias}), and derive the optimal redshift weights for all the relevant parameters, using a specification of the DESI BGS and ELG samples. With the redshift weights shown in Fig. \ref{fig:weight}, we forecast the constraint on the BAO ($\alpha$) and RSD ($\gamma$) parameters, and make a comparison to the case without the redshift weights. We find that with the redshift weights, the FoM of the $\alpha$ and $\gamma$ parameters can be improved by a factor of $2.6$ and $41.6$ respectively for the DESI BGS sample, and $1.9$ and $10.1$ for the ELG sample, which apparently demonstrates the efficacy of the redshift weights, especially for the $\gamma$ parameters. 

We derive the constraint on $\G(z)$ from the $\gamma$ and $\alpha$ parameters, and find that the uncertainty of $\G(z)$ can be reduced by a factor of $2.2\sim12.8$ using the BGS sample at $z \lesssim0.45$, and by a factor of $1.3\sim4.4$ using the DESI ELG sample covering $0.65\lesssim z\lesssim1.65$.

According to our forecast, applying the redshift weights is equivalent to splitting the DESI galaxies into a large number of redshift slices ($\gtrsim40$) for both the BGS and ELG samples, which is impractical due to {the introduced} observational systematics, and to the expense of computational cost. But it is technically straightforward to apply the redshift weights to the galaxy catalogs instead, as performed for the BAO and/or RSD parameters in \cite{RR18,DD18,Zhao18}. {As the weighted catalogs contain the entire galaxy sample, the systematics can be much better controlled, than that for a small fraction of the sample. Admittedly, however, the weights derived from a specific theoretical template based on a fiducial model may be sub-optimal, as the modeled power spectra may be different from the actual ones probed by observations, although the accuracy of the modeling can be improved by iteratively using the observations, it does not bias the result.} 

 The method developed in this work is generally applicable to any redshift survey, including the extended Baryon Oscillation Spectroscopic Survey (eBOSS) \citep{eBOSS1,eBOSS2} \footnote{\url{https://www.sdss.org/surveys/eboss/}}, Prime Focus Spectrograph (PFS) \citep{PFS} \footnote{\url{https://pfs.ipmu.jp/}} and Euclid \citep{Euclid} \footnote{\url{https://www.euclid-ec.org}}.

\acknowledgments

This work is supported by the National Key Basic Research and Development Program of China (No. 2018YFA0404503), National Basic Research Program of China (973 Program) (2015CB857004) and by NSFC Grants 11720101004, 11673025 and 11711530207. This research used resources of the SCIAMA cluster supported by University of Portsmouth.
 
\bibliography{draft}

\appendix

\section{The derivatives of $P_{\ell}$ with respect to the parameters}

In what follows, we show the explicit form of the derivative of $P_{\ell}$ with respect to the parameters, where $\b\equiv b/f$. For brevity, we drop the $z$-dependence of $\ab,\ap,f,b,\gamma, x$, the $k$-dependence of $P_{\rm m}$, and the $k,z$-dependence of $P_0,P_2$ and $P_4$. Only the non-zero derivatives are shown.

\ba && \f{\p\lnPm}{\p \ab}  = -\f{2(35+14\b+3\b^2)(3+\pppk)}{7(15+10\b+\b^2)};   \ \ \f{\p\lnPm}{\p \ap}  = -\f{(35+42\b+15\b^2)(3+\pppk)}{7(15+10\b+\b^2)} \nn \\
&&\f{\p\lnPq}{\p \ab} = -\f{(-7+2\b+\b^2)\pppk+8\b(3+\b)}{2\b(7+3\b)}; \ \ 
\f{\p\lnPq}{\p \ap} = -\f{(7+12\b+5\b^2)\pppk+2\b(9+5\b)}{2\b(7+3\b)} \nn \\
&&\f{\p\lnPh}{\p \ab} = \f{2(11+2\b)\pppk-2(22+19\b)}{11\b}; \ \ 
\f{\p\lnPh}{\p \ap} = \f{-(22+15\b)\pppk+(44+5\b)}{11\b}\nn \\
&&\f{\p\lnPm}{\p f} = \f{2\b(5+3\b)}{f(15+10 \b+3 \b^2)}; \ \ \f{\p\lnPq}{\p f} = \f{7+6\b}{f(7+3\b)}; \ \ \f{\p\lnPh}{\p f} = \f{2}{f}, \nn \\ 
&&\label{eq:A1}\f{\p\lnPm}{\p b} = \f{10\b(3+\b)}{f(15+10 \b+3\b^2)}; \ \ \f{\p\lnPq}{\p b} = \f{7\b}{f(7+3\b)}.\\  \nn\\ \nn\\ 
&&\f{\p\ab}{\p\alpha_0} =1; \  \ \f{\p\ab}{\p\alpha_1}=x; \ \  \f{\p\ab}{\p\alpha_2}=\frac{1}{2}x^2, \nn\\ 
&&\f{\p\ap}{\p\alpha_0} =1; \  \ \f{\p\ap}{\p\alpha_1}=1+2x; \ \  \f{\p\ap}{\p\alpha_2}=x\l(\f{3}{2}x+1\r), \nn\\ 
&&\f{\p\lnf}{\p\alpha_0}=0; \ \ \f{\p\lnf}{\p\alpha_1}=\f{24x}{11}; \ \ \f{\p\lnf}{\p\alpha_2}=\f{6 x(3x+2)}{11}; \ \  \f{\p\lnf}{\p \fzp}=\f{1}{\fzp}, \nn\\
&& \f{\p\lnf}{\p\gamma_0}=3{\rm ln}\l(\f{1+z}{1+\zp}\r)+2{\rm ln}\l(\f{\Hf(\zp)}{\Hf(z)}\r); \ \ \f{\p\lnf}{\p\gamma_1}=\f{\p\lnf}{\p\gamma_0}x; \ \ \f{\p\lnf}{\p\gamma_2}=\f{1}{2}\f{\p\lnf}{\p\gamma_0}x^2,\nn \\ 
&&\label{eq:A2} \f{\p b}{\p\bT}=\f{1}{D(z)}. \ea

\end{document}